\documentclass[12pt,preprint]{aastex}

\usepackage{bm}
\usepackage{xcolor}
\usepackage{amsmath}
\usepackage{lineno}

\colorlet{myc1}{green!20!red!80!}
\definecolor{rkka}{RGB}{219,66,32}

\newcommand{\eb}{\begin{equation}}
\newcommand{\ee}{\end{equation}}

\shorttitle{Exo-Jupiter in $\epsilon$ Eri}
\shortauthors{Makarov et al.}
\begin{document}

\title{Looking for astrometric signals below 20 m/s: A Jupiter-mass planet signature in $\epsilon$ Eri} 
\author{Valeri V. Makarov, Norbert Zacharias, Charles T. Finch}
\affil{United States Naval Observatory, 3450 Massachusetts Ave. NW, Washington, DC 20392-5420, USA}
\email{valeri.makarov@gmail.com}

\begin{abstract}
The USNO ground-based astrometric program URAT-Bright in combination with the Hipparcos mission epoch astrometry provides precise proper motions of a thousand bright stars in the southern hemisphere on a time basis of about 25 years. Small but statistically significant differences between these proper motions and Gaia EDR3 data can reveal long-period exoplanets similar to Jupiter in the nearest star systems. The presence of such a planet orbiting the magnetically active dwarf $\epsilon$ Eri is confirmed from both URAT--Hipparcos--EDR3 data and Hipparcos--EDR3 data
with a corresponding projected velocity of $(+5,+8)$ and $(+6,+13)$ m s$^{-1}$,
respectively. These signals are formally significant at a 0.989 and 1.0 confidence. We conclude that the newest astrometric results confirm the existence of a long-period exoplanet orbiting $\epsilon$ Eri, which was marginally detected from precision radial velocity measurements some 20 years ago.

\end{abstract}


\section{Introduction}
\label{intr.sec}
The rapidly growing compendium of known exoplanets is strongly biased toward
short-period systems, whose detection with the most productive photometric transit and spectroscopic radial velocity methods benefits from stronger and more frequent signals. The astrometric ``$\Delta\mu$" method of detecting dim companions by the reflex motion effect on observed proper motions \citep{2005AJ....129.2420M} is presently a viable and economic alternative to decade-long radial velocity campaigns owing to the increased accuracy of short-term proper motions in Gaia EDR3 \citep{lin}. With the covered mission duration shorter than the orbital period, the mean proper motion includes most of the projected reflex orbital motion. A long-term proper motion derived from two absolute celestial positions separated by a time interval that is much longer than the orbital period is likely to be less affected by the reflex orbital motion. The difference between the two proper motion vectors betray the presence of a companion with a period of at least several years.

The required sensitivity of the $\Delta\mu$-method to detect a Jupiter analog orbiting a Sun-like host is
13 m s$^{-1}$, or $0.27$ mas yr$^{-1}$ at a distance of 10 pc. The most recent Gaia data release \citep[EDR3,][]{gai} has achieved this or better performance for many nearby stars. Obtaining long-term proper motions at the required level of accuracy is more problematic. We use two sources of data in this study: 
a combined Hipparcos-URAT-UBAD astrometric solution
(HUU) obtained at USNO within the framework of the USNO Bright Star Catalog
program and ad hoc proper motions directly computed from the mean positions in the Hipparcos main catalog \citep{esa} and in the Gaia EDR3 catalog. The HUU solution is based on the Hipparcos Individual
Astrometry Data (HIAD) and epoch measurements recently collected with ground-based telescopes, including the USNO Robotic Astrometric Telescope \citep[URAT,][]{zac} in La Serena, Chile. The average precision of the
URAT-HIAD proper motions with a timeline of 22--26 years is approximately 0.20 mas yr$^{-1}$ per coordinate.

\section{The planet of $\epsilon$ Eri}
\label{star.sec}

$\epsilon$ Eri = HIP 16537 is a nearby magnetically active K2V dwarf with a Gaia EDR3 parallax of $\varpi=310.58\pm 0.14$ mas. A giant long-period planet was suggested by \citet{hat} based on re-analysis of previously published radial velocity measurements taken with different telescopes and instruments over a period of about 19 years. The reported parameters were orbital period $P=6.7$
yr, eccentricity $e=0.6$, velocity amplitude $K=19$ m s$^{-1}$, and projected mass $M\,\sin i=0.86 M_{\rm J}$. 
Doubts about the validity of this detection lingered for a while because of a high level of noise and jitter in the spectroscopic data. A planet with a period
$P=7.37\pm 0.07$ yr of similar mass but a nearly circular orbit was confirmed by \citet{maw}. The prominent dusty debris disk of $\epsilon$ Eri has an inner radius of
$\sim 30$ au \citep{gre} possibly caused by another giant planet at this distance \citep{hat}--- however, no such object has been found by direct imaging \citep{pat}.

\section{$\Delta\mu$ and its statistical significance}

The equatorial system proper motion from EDR3 at the mean epoch of 2016 is
$(-974.758\pm 0.160, 20.876\pm0.120)$ mas yr$^{-1}$. The values are within one formal error of the Gaia DR2 proper motions, which were used by \citet{ker} to detect the signal from the proposed exo-Jupiter, with a null result. The HUU proper motions are $(-975.08\pm 0.17, 20.33\pm0.18)$ mas yr$^{-1}$.
The proper motion difference vector EDR3$-$HUU of $\boldsymbol{d}=\Delta\mu=(0.322, 0.546)$ mas yr$^{-1}$ corresponds to a tangential velocity vector 
$(+5, +8)$ m s$^{-1}$. This tiny signal is statistically significant as follows from a computation of the expected error. The full covariance matrix $\boldsymbol{C}$ comprises the covariance of the HUU vector (a 2 by 2 matrix) and the corresponding Gaia EDR3 covariance, which is specified in the catalog. The squared SNR value is
\eb
\delta^2=\boldsymbol{d}'\,\boldsymbol{C}^{-1}\,\boldsymbol{d},
\ee
which is expected to be $\chi^2(2)$-distributed. We obtain $C_{11}=0.0580$, $C_{22}=0.0468$, $C_{12}=-0.0053$, $\delta^2=8.94$, and
a confidence level of 0.989 for this detection from the cumulative distribution function CDF[$\chi^2(2)$].

This result can be verified by computing a
long-term proper motion directly from the difference in the mean positions in Hipparcos (mean epoch
1991.25) and Gaia EDR3 (epoch 2016). The full covariance matrix in this case
includes two parts, the sum of the position covariance matrices for Hipparcos and EDR3 divided by $24.75^2$ (the squared epoch difference) and the covariance
matrix of the EDR3 proper motion. Here we ignore the small contribution from the covariances between EDR3 position and
proper motion. Repeating the steps
for EDR3$-$HUU difference, we obtain a $\Delta\mu$ of $(0.410, 0.849)$ mas yr$^{-1}$, which corresponds to a velocity signal of $(6,\,13)$ m s$^{-1}$.
Because of the smaller variances and larger magnitude, the $\delta^2$ value comes up to 68.0 and the confidence is 1.

\begin{figure}[htbp]
  \centering
  \plotone{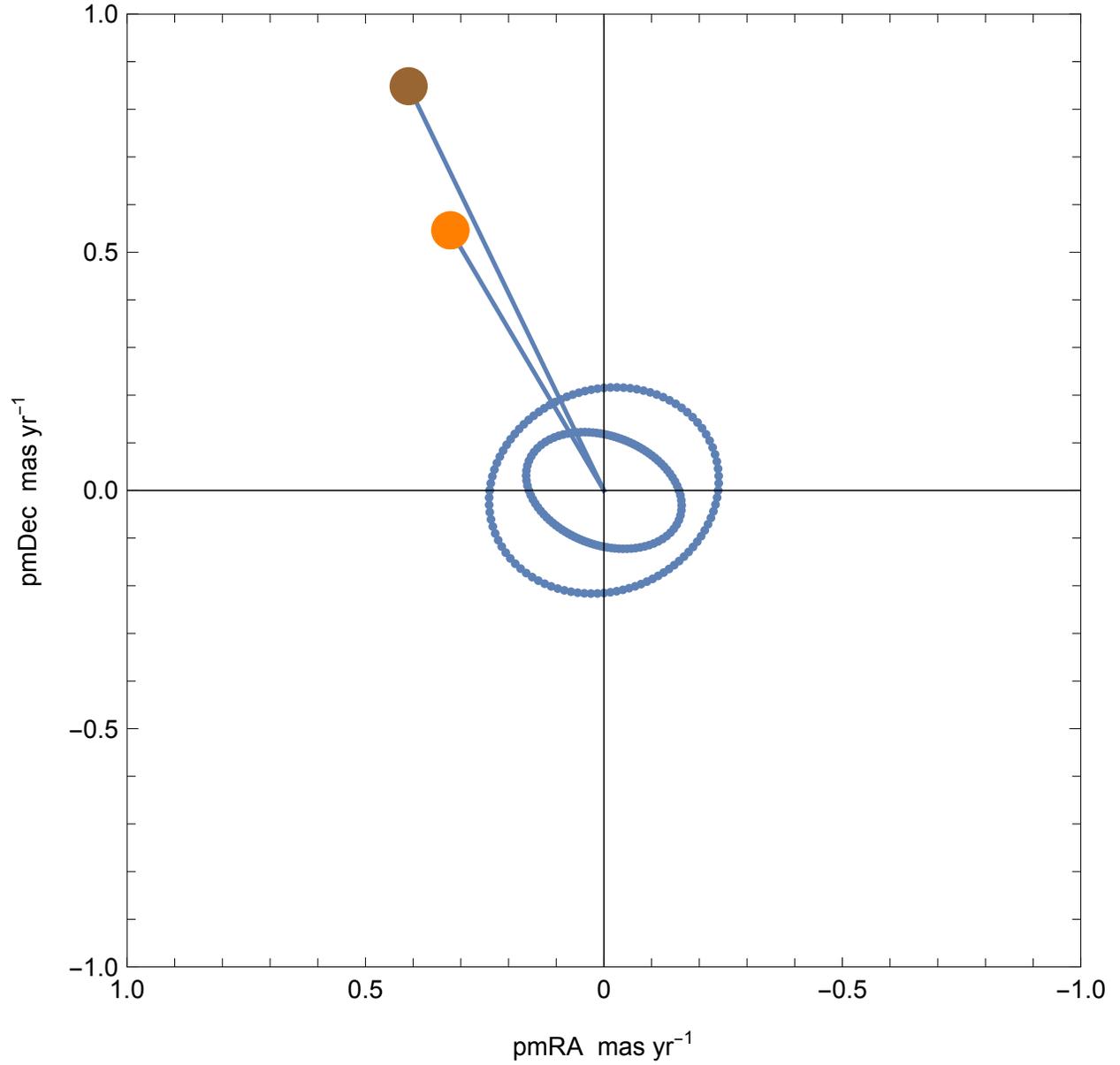}
\caption{Proper motion difference vectors EDR3$-$HUU and EDR3$-$HG for $\epsilon$ Eri marked with large brown and orange circles, respectively, 
and the corresponding error ellipses in the equatorial sky tangent plane. The larger ellipse is for the EDR3$-$HUU data. \label{ellipse.fig}}
\end{figure}

Fig. \ref{ellipse.fig} shows the error ellipses and the actual $\Delta\mu$ vectors computed from the two data sets. The results are consistent between the two methods, taking into account that some differences can be caused by a small offset in the timelines of HUU and Hipparcos--Gaia derivation. The formal confidence in this result is close to unity. 

At this level of accuracy, apparent acceleration of nearby stars can be
caused by the radial velocity component of the space velocity vector, the effect known as perspective acceleration. This acceleration is always aligned
with (or opposite to) the proper motion vector on the sky, which is not
the case for $\epsilon$ Eri.

\section{Conclusions}
\label{con.sec}
Using two somewhat different techniques and data sets, we derived the proper motion difference vectors between
long-term URAT-Bright data and the short-term Gaia EDR3 proper motions.
Both techniques indicate the presence of a small astrometric signal at high formal confidence levels.
The results are consistent with the previously reported planet epsEri-b of approximately Jupiter mass and a period of several years. Ultimately, astrometric position measurements obtained with Gaia should reveal the reflex orbit of $\epsilon$ Eri and provide its accurate ephemerides.

\end{document}